\documentclass[structabstract,letter,a4paper]{aa} 
\usepackage{graphicx}
\usepackage{txfonts}
\usepackage{natbib}
\bibpunct{(}{)}{;}{a}{}{,}
\def\rpa{1.338$\pm$0.013}
\def\rpb{1.338$\pm$0.016}
\def\tchisq{1.58}
\def\tsig{0.0060}
\def\dfa{$-$0.241$\pm$0.021}
\def\dfb{$-$0.241$\pm$0.043}
\def\dfc{$-$0.190$^{+0.034}_{-0.037}$}
\def\dfd{$-$0.233$\pm$0.020}

\def\Tb{2040$\pm$185}

\def\ecosom{$-$0.0066$\pm$0.0021}
\def\sesig{0.0038}
\def\sechisq{1.72}
\def\sechisqb{1.39}
\def\sechisqd{1.77}
\begin{document}
   \title{Ground-based K-band detection of thermal emission from the exoplanet TrES-3b}
\titlerunning{Ground-based detection of emission from TrES-3b}
   \author{E.J.W. de Mooij\inst{1}
           \and
           I.A.G. Snellen \inst{1}
          }
    \institute{Leiden University, Postbus 9513, 2300 RA, Leiden, 
               The Netherlands;    \email{demooij@strw.leidenuniv.nl}
               }
    \date{  }

\abstract
{Secondary eclipse measurements of transiting extrasolar planets with the 
Spitzer Space Telescope have yielded several direct detections of thermal
exoplanet light. Since Spitzer operates at wavelengths longward of 
3.6$\mu$m, arguably one of the most interesting parts of the planet spectrum 
(from 1 to 3 $\mu$m) is inaccessible with this satellite. This region is at the 
peak of the planet's spectral energy distribution and is also the regime where 
molecular absorption bands can significantly influence the measured emission.}
{So far, 2.2$\mu$m K-band secondary eclipse measurements, which are possible 
from the ground, have not yet lead to secure detections. The aim of this paper 
is to measure the secondary eclipse of the very hot Jupiter TrES-3b in K-band, 
and in addition to observe its transit, to obtain an accurate planet radius
in the near infrared.}
{We have used the William Herschell Telescope (WHT) to observe the secondary 
eclipse, and the United Kingdom Infrared Telescope (UKIRT) to observe the 
transit of TrES-3b. Both observations involved significant defocusing of the 
telescope, aimed to produce high-cadence time series of several thousand frames 
at high efficiency, with the starlight spread out over many pixels.}
{We detect the secondary eclipse of TrES-3b with a depth of \dfb\% 
($\sim$6$\sigma$).
This corresponds to a day-side brightness temperature of T$_B$(2.2$\mu$m)=\Tb K,
which is consistent with current models of the physical properties of this
planet's upper atmosphere. The centre of the eclipse seems slightly offset from 
phase $\phi$=0.5 by $\Delta\phi$=$-$0.0042$\pm$0.0027, which could indicate that the orbit of
TrES-3b is non-circular. Analysis of the transit data shows that TrES-3b has a 
near-infrared radius of \rpb R$_{Jup}$, showing no significant deviation from 
optical measurements.}
{}


\maketitle

\section{Introduction}
During the secondary eclipse of an extrasolar planet (the moment the planet 
moves behind its host star), the contribution from the direct planet light can
be measured. Until now such measurements have been the domain of the Spitzer 
Space Telescope, which have led to the first detections of thermal emission 
from extrasolar planets~\citep{charbonneauetal05, demingetal05}. Until now a 
handful of planets has been observed at several wavelengths between 3.6$\mu$m 
and 24$\mu$m, with one of the highlights so far being the measurement of the 
day/night-side temperature contrast of the exoplanet HD189733b~\citep{knutsonetal07b}.

Unfortunately, a very interesting part of the broad spectral energy 
distribution of hot Jupiters between 1 and 3 $\mu$m is short-ward of the 
observing window of Spitzer. This region contains the overall peak of the 
planet's emission spectrum, and can also be strongly influenced by molecular 
bands (e.g. from water, methane and CO) in the planets' upper atmosphere.
However, ground-based observations, which could access this spectral region, 
have thus far proven very challenging. Several methods have been tried. 
\cite{richardsonetal03b} used a differential spectroscopic method, searching for
the wavelength dependent shape of the planet spectrum, but no eclipse was 
detected. In addition, observations of the secondary eclipse of TrES-1b in 
L-band with the NIRI spectrograph on Gemini North by~\cite{knutsonetal07b} only 
resulted in an upper limit.
 
Photometric attempts to detect the secondary eclipse of extrasolar planets
have been presented by~\cite{snellen05} and~\cite{Snellen_covino07}. Two partial
eclipses of HD209458b were observed in K-band by~\cite{snellen05}, while the 
telescope was significantly defocused to avoid saturation of the array by the
bright star. Although milli-magnitude precisions were reached, nothing was 
detected, probably because of the lack of baseline on both sides of the
eclipse~\citep{snellen05}.
Subsequently, OGLE-TR-113b was observed~\citep{Snellen_covino07}, which has the 
great advantage that it has several other stars in its field. It was shown that 
randomly offsetting the  telescope randomises the photometric errors down to 
0.1-0.2\% per hour. It resulted in a tentative detection of its secondary 
eclipse of 0.17$\pm$0.05\%.

In this letter we present our results for K-band photometry of the 
transit and secondary eclipse of the planet TrES-3b~\citep{odonovan2007}. 
This planet is significantly more suitable for secondary eclipse photometry 
than previous targets, since it is in a very close orbit (P$\sim$1.31 days) and
its radius is significantly inflated (R$\sim$1.3R$_{Jup}$). Furthermore there is
a nearby reference star, which can be used for differential photometry. 
In section~\ref{sec:obs_dr} we present our observations and data 
reduction. In section~\ref{sec:results} we present and discuss our results.

\section{Observations, data reduction and analysis}\label{sec:obs_dr}
\subsection{The transit of TrES-3b}
\begin{figure}
\centering
\includegraphics[width=8cm]{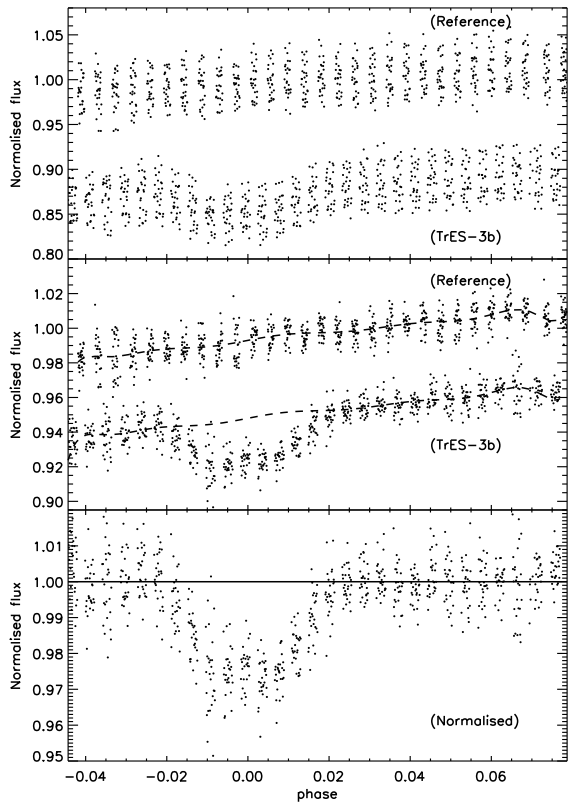}
\caption{K-band lightcurves from the UKIRT transit observations. 
The top panel shows the lightcurve of the reference star and TrES-3b 
before correcting for the positional dependence of the flux. The middle panel 
shows the fluxes of both stars corrected for this effect. 
The dashed line is a polynomial fit to the flux of the reference star. 
The bottom panel shows the final corrected unbinned lightcurve of TrES-3b.}
\label{fig:UFTI_lc_raw}
\end{figure}
\begin{figure}
\centering
\includegraphics[width=8cm]{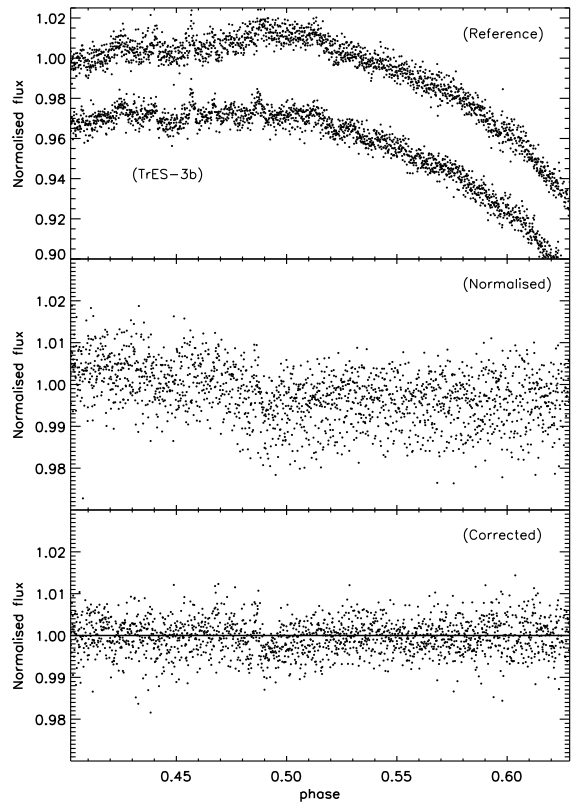}
\caption{K-band lightcurves from the WHT secondary eclipse observations. 
The top panel shows the raw lightcurve for the reference star and TrES-3b. The 
middle panel shows the lightcurve of TrES-3b divided by that of the reference 
star. The bottom panel shows the final, fully corrected, lightcurve.}
\label{fig:LIRIS_lc_raw}
\end{figure}
We have observed the transit of the exoplanet TrES-3b with the United Kingdom 
Infrared Telescope (UKIRT) using its Fast Track 
Imager~\citep[UFTI;][]{rocheetal03} on June 20, 2008. The observations, carried
out in queue scheduling, lasted for almost 4 hours, starting $\sim$0.6 hour 
before ingress and ending $\sim$1.7 hours after egress.
The field of view of UFTI is 93$\times$93 arcsec, with a pixel scale of 0.091 
arcsec/pixel. Since the distance between TrES-3 (K=10.61) and the nearby 
reference star (2MASS J175225.15+373422.1, K$\sim$9.77) is $\sim$4 arcminutes,
we were forced to alternate observations between the target and the reference,
in order to correct for time dependent atmospheric and instrumental effects.
To reduce overheads from detector read-out, the detector was windowed to a 
single quadrant (512$\times$512 pixels, 46.6$\times$46.6 arcsec) which increased
the cycle speed. 
During each sequence we repeated a nine point dither pattern three times before 
switching to the other star. 
The exposure time was 5 seconds for the TrES-3, and 2 seconds for the reference 
star. With a typical overhead of 5 seconds per frame, the time required to 
complete an entire target-reference sequence was $\sim$8 minutes.

The telescope was significantly defocused to 1) reduce intrapixel variations, 2)
minimise the influence of flatfield inaccuracies, and 3) prevent the count 
levels from reaching the non-linear range of the detector. 
Due to the fact that the telescope optics had not yet been realigned after the
recent switch to the Cassegrain focus, the defocused telescope produced a 
significantly asymmetric PSF, which could be covered by a circular aperture
of $\sim$40 pixels ($\sim$4") in radius.

The data was flatfielded using a sky flat constructed from a series of frames 
taken just before the start of the observations of TrES-3.
No dark subtraction was performed, because any dark-current was seen as a
contribution to the sky background and removed together with the sky. 
The positions of the hot and cold pixels were determined from both the 
flat field and separate dark frames. These pixels were replaced by the values of
3rd order polynomial surfaces fitted to the 7$\times$7 pixels surrounding these 
pixels.

To determine the sky background, first all stars in the field were masked 
using circular masks of 30 pixels. A median sky value for the masked image was 
calculated and subtracted in order to remove temporal fluctuations in the 
background. The final sky level was determined by averaging the 27 masked 
frames from one cycle, and removed from the images.

 Subsequently, aperture photometry was performed using the APER procedure 
from the IDL Astronomy User's Library\footnote{http://idlastro.gsfc.nasa.gov\/} 
with an aperture radius of 42 pixels. 
Any possible residual sky fluctuations were corrected for by measuring the 
background value in a 42 to 70 pixel annulus around the object.

 The results from the aperture photometry can be found in the top panel of 
Figure~\ref{fig:UFTI_lc_raw}. Both the flux from the target and the reference
star was found to be a function of dither position, therefore the data were 
normalised for each dither position separately. These corrected lightcurves are
shown in the middle panel of Figure~\ref{fig:UFTI_lc_raw}. 
To remove time dependent effects, we fitted a high order polynomial to the 
binned lightcurve of the reference star. This fit was used to correct the 
lightcurve of TrES-3. The final unbinned lightcurve for TrES-3 is shown in the 
bottom panel of Figure~\ref{fig:UFTI_lc_raw}.

\subsection{The secondary eclipse of TrES-3b}
 On July 3-4, 2008 the secondary eclipse of TrES-3 was observed with the 
Long-slit Intermediate Resolution Infrared Spectrograph~\cite[LIRIS;][]{LIRIS02} at the William
Herschell Telescope (WHT) on La Palma, with the observations lasting for the 
entire night, starting about $\sim$3 hours before, and ending $\sim$4 hours 
after the expected centre of the secondary eclipse (at 1:31UT on July 4th).
We used the full 1024$\times$1024 pixel array for these observations, which 
combined with the pixel size of LIRIS of 0.25"/pixel, yields a field of view 
of 256$\times$256 arcseconds. This larger field of view enabled us to place 
both TrES-3 and the reference star simultaneously on the detector, making these
observations $\sim$80\% more efficient relative to those obtained with UKIRT.
We used a 9 point dither pattern, including small random offsets from the 
nominal positions of each dither point.
We used an exposure time of 7.5 seconds, with typical overheads of 7 seconds 
per frame. 

As for the UKIRT observations, we defocused the telescope, which resulted in a 
ringshaped PSF with a radius of about 4 pixels.

 Since the LIRIS detector is non linear at high count levels, we applied the 
following empirical non-linearity correction to the data before flat fielding:
\begin{eqnarray}
F_{true}&=&F_{meas}\cdot\frac{1}{1-c\cdot F_{meas}^6} 
\end{eqnarray}
with $F_{true}$ the flux after correction, $F_{meas}$ the measured flux and 
$c=6\cdot10^{-29}$ a constant describing the strength of the correction term. 
The value of $c$ was determined by minimising the noise of the reference 
star over the entire night, and comparing this value to non-linearity 
measurements taken by the LIRIS team\footnote{http://www.ing.iac.es/Astronomy/instruments/liris/liris\_detector.html}.

All frames were corrected for crosstalk between the quadrants, which
was found to be present in a co-add of all the reduced science frames at a 
level of $10^{-5}$ of the total flux along a row of a quadrant.
This correction was done on a row by row basis. A flat field was created from a 
series of domeflats taken with the domelight both on and off. These were 
subtracted from each other (to eliminate structure from the emission by the dome 
and telescope) and averaged.

The small number of hot and cold pixels in the image were replaced by a 
median of the neighbouring pixels in a 3$\times$3 grid. 
The sky was determined from the combination of the 9 images in each dither
sequence, for which all the stars were masked out. We used a circular 
aperture with a radius of 13 pixels to determine the stellar flux 
and an annulus of 30 to 90 pixels to determine the level of possible residual
flux in the background after sky-subtraction.

The individual raw lightcurves of TrES-3b and the reference star are plotted in 
the top panel of Figure~\ref{fig:LIRIS_lc_raw}. Before further analysis we 
divided the lightcurve of TrES-3 by that of the reference star, shown in the 
middle panel of Figure~\ref{fig:LIRIS_lc_raw}. The relative fluxes of the two 
stars are correlated with their positions on the detector. These effects are 
corrected for by fitting linear functions between the x,y-positions and the 
relative flux for data outside of the expected eclipse. These fits are then 
applied to the whole dataset. In a similar way a small dependence on
airmass was removed. The corrected unbinned lightcurve of TrES-3b can be found 
in the lower panel of Figure~\ref{fig:LIRIS_lc_raw}.

\section{Results and discussion}\label{sec:results}
\subsection{The transit of TrES-3b}
\begin{figure}
\centering
\includegraphics[width=8cm]{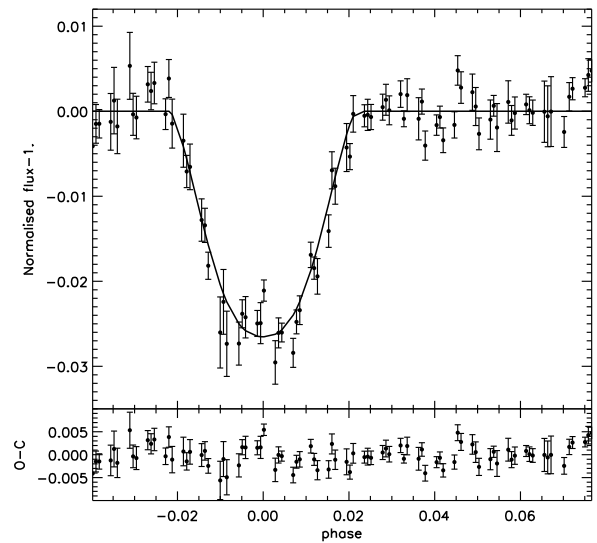}
\caption{The 9-point binned lightcurve of the transit of TrES-3b. The solid line shows the best
fitting model with R$_P$=\rpb R$_{Jup}$. The bottom panel shows the residuals 
after subtracting this model fit.}
\label{fig:UFTI_lc_red}
\end{figure}
\begin{figure}
\centering
\includegraphics[width=8cm]{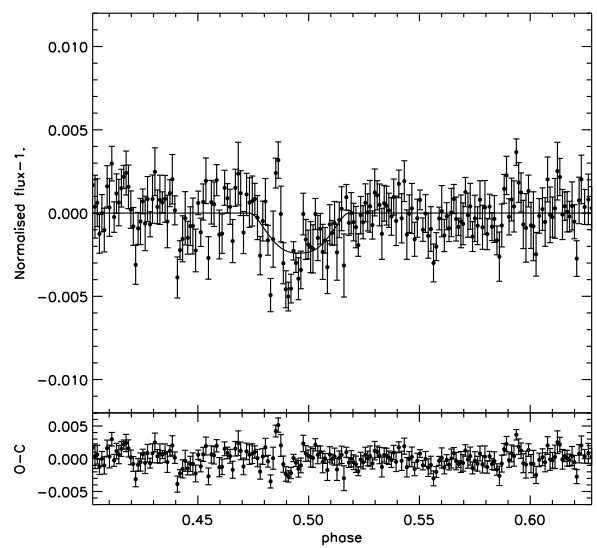}
\caption{Final, 9-point binned lightcurve of the secondary eclipse. The solid line
is the best fitting model with $\Delta F$=\dfb\% and an offset from phase 0.5 of 
$\Delta \phi $=$-$0.0041$\pm$0.0018. The bottom panel shows the residuals after 
subtracting the best fitting model.}
\label{fig:LIRIS_lc_fin}
\end{figure}
The final transit lightcurve binned per 9-point dither cycle is shown in 
Figure~\ref{fig:UFTI_lc_red}. The RMS-noise level out of transit is \tsig, 
while the theoretical noise level for these observations is 0.0029, 
which is actually dominated by the detector's readnoise of 26e$^-$/pixel, due 
to the large aperture used.

 Since a dependence of the transit depth on wavelength can reveal  
characteristics of the planet atmosphere, we fitted the radius of TrES-3b, 
keeping the parameters of the host star and the planet's orbit 
(mean stellar density, orbital inclination and ephemeris) fixed with 
respect to the values of~\cite{sozzettietal08}.
For the limb darkening we used parameters from~\cite{claret2000}, assuming a 
temperature of $T=5750$K for the host star of TrES-3b, a metalicity of 
log[Fe/H]$=-0.2$ and a surface gravity of log(g)$=4.5$. 
The transit was modelled using the IDL procedures from~\cite{mandel2002}, 
resulting in a radius of \rpa R$_{Jup}$ at $\chi^2/\nu=$\tchisq. The 
reduced $\chi^2$ is significantly larger than unity. Since we fitted to the 
binned lightcurve, this indicates that the noise does not scale with $\sqrt{N}$,
meaning that we have an additional, unknown, source of correlated noise. To 
obtain a better estimate of the uncertainty in the planet radius, we forced
$\chi^2/\nu=1$ by scaling up our errors in the binned lightcurve uniformly. In 
this way we obtain R$_P$=\rpb R$_{Jup}$. The K-band radius shows no significant 
deviation from the optical radius of 1.336$_{-0.037}^{+0.031}$R$_{Jup}$, as 
measured by~\cite{sozzettietal08}. 
The uncertainty in our K-band radius appears smaller, but this is solely due to 
the fact that we keep the parameters for the host star and the planet's orbit 
fixed.

We show that we can obtain the K-band radius of an exoplanet at a precision of
$\sim$1\%. This could be further improved by observing the reference star and 
target at the same time both because of the increase in the number of frames 
during transit and because rapid atmospheric fluctuations can be corrected for.

\subsection{The secondary eclipse of TrES-3b}
The fully corrected lightcurve binned by one 9 point dither cycle is shown in 
Figure~\ref{fig:LIRIS_lc_fin}. Before binning, and after subtracting the best 
fitting model (see below), we clipped the lightcurve at $\pm$0.01, removing 
43 of the 1800 points (2.3\%), of which 3 points (0.9\%) during the eclipse.
 The  binned lightcurve was fitted using the~\cite{mandel2002} 
procedures, with all the parameters fixed to the transit model, except the 
centre and the depth of the eclipse (of course without limb darkening).

The best fitted lightcurve is over-plotted in the top panel of 
Figure~\ref{fig:LIRIS_lc_fin}, with the bottom panel showing the residuals.
The out of eclipse RMS noise is \sesig, while the expectation from theoretical 
noise statistics is 0.0015. The dominant source of noise in these noise 
calculations is the sky contribution. The total flux from the sky within the 
aperture is, even for the brighter reference star, $\sim$50\% higher than the stellar flux.

The fitted depth of the secondary eclipse is $\Delta F=$\dfa\%,
with a reduced chi-squared of $\chi^2/\nu$=\sechisq. 
As for the transit data, the reduced $\chi^2$ for the fit to the binned 
lightcurve is greater than unity, indicating that we have some extra systematic 
noise. We also find a timing difference in phase of   $\Delta \phi $=$-$0.0041$\pm$0.0013 (at $\chi^2/\nu$=\sechisq).
The offset in the timing is intriguing, because it may indicate that the orbit 
of TrES-3b is not circular but slightly eccentric, with ecos$\omega$
=\ecosom. Although the formal significance of the offset is $\sim$3$\sigma$, 
the residual systematic noise in the lightcurve decreases its significance,
in particular due to several points around $\phi=0.49$. We also refitted the data with the centre of the eclipse fixed to $\phi$=0.5, and obtained in this case a depth of \dfd $ $ at $\chi^2/\nu$=\sechisqd.

To test the influence of the more deviated points around $\phi$=0.49, we 
repeated the fit while excluding them. The depth subsequently becomes \dfc, 
with an offset in phase of $\Delta\phi=$$-$0.0036$^{+0.0022}_{-0.0020}$ ($\chi^2/\nu$=\sechisqb). 
Although the significance of the fit is reduced, we still detect the secondary 
eclipse at a $6\sigma$-level.

 Because it is clear that red noise dominates the uncertainties in our measurements,
we tried to characterise the uncertainty using the "residual permutation" 
bootstrap method as used by~\cite{winnetal08}. The best fitting model was subtracted from the data, after which the residuals were shifted between 1 and 1800 points and added to the model light curve. These light curves were refitted for each individual shift. The resulting distributions of parameter values represent better estimates of the real uncertainties in the data. 
 With this analysis we find an uncertainty of 4.3$\cdot$10$^{-4}$ in the eclipse depth 
and 2.7$\cdot$10$^{-3}$ for the offset in phase, which we use in the remainder of the paper.

Assuming a blackbody model for the thermal emission spectrum of TrES-3b, we can
estimate the day-side brightness temperature in the K-band. 
For the stellar temperature, we use T$_{\rm{eff}}$=5650$\pm$75 K~\citep{sozzettietal08}. 
From the fit to the eclipse depth using the full lightcurve we obtain T$_B$=\Tb K (90\% confidence interval).

Due to the high incident stellar flux, \cite{fortneyetal08} classify this
planet as belonging to the pM class of planets. This class is expected to have
an inversion layer in their upper atmospheres and to show a large day/night 
contrast. Comparing the depth of the secondary eclipse with their model 
(their Figure 14), we see that our results are indeed consistent with this
model. However, since we have measured the brightness temperature of TrES-3b at 
only single wavelength, Spitzer measurements at longer wavelengths are 
needed to confirm the presence of an inversion layer and thus classify TrES-3b 
as a pM planet.

\section{Conclusions}
We have detected the secondary eclipse of the exoplanet TrES-3b in K-band at a
level of \dfb\%, indicating a brightness temperature of T$_B$(2.2$\mu$m)=\Tb K.
The eclipse timing shows a small offset from $\phi$=0.5, which would imply a 
non-circular orbit, but this needs to be confirmed by Spitzer observations.
Also the radius of TrES-3b is determined in K-band to be \rpb R$_{Jup}$, showing
no deviations from optical measurements.

\begin{acknowledgements}
We are grateful to the UKIRT observers and the staff of both the UKIRT and 
WHT telescopes.
The United Kingdom Infrared Telescope is operated by the Joint Astronomy Centre 
on behalf of the Science and Technology Facilities Council of the U.K. 
The William Herschel Telescope is operated on the island of La Palma 
by the Isaac Newton Group in the Spanish Observatorio del Roque de los 
Muchachos of the Instituto de Astrof\'isica de Canarias.

\end{acknowledgements}
\bibliographystyle{aa} 
\bibliography{1239} 
\end{document}